\documentstyle[preprint,prl,aps]{revtex}
\tightenlines
\begin{document}
\title{Fokker-Planck equation with variable diffusion coefficient in the
Stratonovich approach}
\author{Kwok Sau Fa}
\address{Departamento de F\'{i}sica, \ Universidade \ Estadual de Maring\'{a}, \ \\
Av. Colombo 5790, 87020-900, Maring\'{a}-PR, Brazil}
\maketitle

\begin{abstract}
We consider the Langevin equation with multiplicative noise term which
depends on time and space. The corresponding Fokker-Planck equation in
Stratonovich approach is investigated. Its formal solution is obtained for
an arbitrary multiplicative noise term given by $g(x,t)=D(x)T(t)$, and the
behaviors of probability distributions, for some specific functions of $D(x)$%
, are analyzed. In particular, for $D(x)\sim \left| x\right| ^{-\theta /2}$,
the physical solutions for the probability distribution in the Ito,
Stratonovich and postpoint discretization approaches can be obtained and
analyzed.
\end{abstract}

\newpage

In these last decades, anomalous diffusion properties have been extensively
investigated by several approaches in order to model different kinds of
probability distributions such as long-range spatial or temporal
correlations \cite{klafter}. For instance, the well-known cases are the
Langevin and the corresponding Fokker-Planck equation, and the master
equation. The other ones we could mention are the generalized Langevin
equations \cite{mura}, the generalized Fokker-Planck equation with memory
effect \cite{risken}, generalized thermostatistics \cite{tsallis},
generalized master equations \cite{kenkre}, continuous time random walk \cite
{klafter2}, and fractional equations \cite{klafter}. These approaches have
been used to describe numerous systems in several contexts such as physics,
hydrology, chemistry and biology.

The well-established property of the normal diffusion described by the
Gaussian distribution can be obtained by the usual Fokker-Planck equation
with a constant diffusion coefficient (without the drift term). Anomalous
diffusion regimes can also be obtained by the usual Fokker-Planck equation,
however, they arise from variable diffusion coefficient which depends on
time and/or space. On the other hand, in the view of Langevin approach it is
associated with a multiplicative noise term. In other approaches such as the
generalized Fokker-Planck equation (nonlinear) and fractional equations,
they can describe anomalous diffusion regimes with a constant diffusion
coefficient.

In this work, we investigate the Fokker-Planck equation with variable
diffusion coefficient in time and space, in the Stratonovich approach. We
show that for a multiplicative noise term separable in time and space, $%
g(x,t)=D(x)T(t)$, we can obtain a formal solution for the probability
distribution. We also analyze the behaviors of probability distributions for
some specific functions of $D(x)$ which can manifest interesting properties
such as non-Gaussian distribution, combination of behaviors like Gaussian
(for small distance) and exponential (for large distance), and combination
of behaviors like Gaussian (for small distance) and power law decay for long
distance. Also, we can obtain many bimodal distributions for different forms
of $D(x)$.

Now, we consider the following Langevin equation

\begin{equation}
\dot{\xi _{i}}=h_{i}(\xi ,t)+g_{ij}(\xi ,t)\Gamma _{j}(t)\text{ ,}
\label{eq1}
\end{equation}
where $\xi _{i}$ is a stochastic variable and $\Gamma _{i}(t)$ is the
Langevin force. We assume that the averages $\left\langle \Gamma
_{i}(t)\right\rangle =0$ and $\left\langle \Gamma _{i}(t)\Gamma _{j}(%
\overline{t})\right\rangle =2\delta _{ij}\delta (t-\overline{t})$ \cite
{risken}. For $h=0$ and $g=D$, Eq.(\ref{eq1}) describes the Wiener process
and the corresponding probability distribution is described by a Gaussian
function. In the case of $g(\xi ,t)$ variable, some specific functions have
been employed to study, for instance, turbulent flows $(g(x,t)\sim
\left| x\right|^{a}t^{b})$ \cite{richar}, relativistic Brownian motion $(h(p,t)\sim p$ and 
$g(p,t)\sim \left( 1+(p/mc)^{2}\right) ^{1/4})$ \cite{hanggi} and classical
harmonic oscillator \cite{gitter}. By applying the Stratonovich approach in
a one-dimensional space \cite{risken}, we obtain the following dynamical
equation for the probability distribution

\begin{equation}
\frac{\partial W(x,t)}{\partial t}=-\frac{\partial \left[ h(x,t)W(x,t)\right]
}{\partial x}+\frac{\partial }{\partial x}\left[ g(x,t)\frac{\partial \left(
g(x,t)W(x,t)\right) }{\partial x}\right] \text{ .}  \label{eq2}
\end{equation}

Hereafter we consider the drift term equal to zero ($h=0$). We also consider
that the multiplicative noise term $g(x,t)$ is separable in time and space, $g(x,t)=D(x)T(x)$.
Then, we can rewrite Eq. (\ref{eq2}) as the following manner

\begin{equation}
\frac{\partial \rho }{\partial \overline{t}}=\frac{\partial ^{2}\rho }{%
\partial \overline{x}^{2}}  \label{eq3}
\end{equation}
with

\begin{equation}
\rho (x,t)=D(x)W(x,t)\text{ ,}  \label{eq4}
\end{equation}

\begin{equation}
\frac{d\overline{t}}{dt}=T^{2}(t)  \label{eq5}
\end{equation}
and

\begin{equation}
\frac{d\overline{x}}{dx}=\frac{1}{D(x)}\text{ .}  \label{eq6}
\end{equation}
Eq. (\ref{eq3}) has the following formal solution,

\begin{equation}
\rho (\overline{x},\overline{t})=A\frac{\exp \left[ -\frac{\overline{x}^{2}}{%
4\overline{t}}\right] }{\sqrt{\overline{t}}}  \label{eq7}
\end{equation}
where $A$ is a normalization factor. We note that for $D(x)=\sqrt{D}$ and $%
T(t)=1$ we recover the Wiener process.

We can now explore some spatial features of the probability distribution of
solution (\ref{eq4}), for some specific forms of $D(x)$:

First case - \ We consider

\begin{equation}
\overline{x}=\frac{x}{\sqrt{D}\left( 1+a\left| x\right| ^{c}\right) ^{b}}%
\text{ .}  \label{eq8}
\end{equation}
where $a$ is a positive real number. From Eqs. (\ref{eq4}), (\ref{eq6}) and (%
\ref{eq7}) we obtain

\begin{equation}
W(x,t)=C_{1}\frac{\left[ 1+a(1-bc)\left| x\right| ^{c}\right] \exp \left[ -%
\frac{x^{2}}{4D\overline{t}\left( 1+a\left| x\right| ^{c}\right) ^{2b}}%
\right] }{\sqrt{\overline{t}}\left( 1+a\left| x\right| ^{c}\right) ^{1+b}}%
\text{ .}  \label{eq9}
\end{equation}
where $bc<1$ in order to maintain $W(x,t)$ positive. For $a=0$ or $b=0$
 we recover the Wiener process.

In particular, for $c=1/b$, we have

\begin{equation}
W(x,t)=C_{2}\frac{\exp \left[ -\frac{x^{2}}{4D\overline{t}\left( 1+a\left|
x\right| ^{1/b}\right) ^{2b}}\right] }{\sqrt{\overline{t}}\left( 1+a\left|
x\right| ^{1/b}\right) ^{1+b}} \text{  .}  \label{eq10}
\end{equation}
In this process, the behavior, for small $a$ and $x,$ is like a Gaussian
function, whereas for large distance, the exponential term converges to a
constant value. Therefore, for large distance, the dominant term is the
multiplicative factor $1/\left( 1+a\left| x\right| ^{1/b}\right) ^{1+b}$
which approximates to the asymptotic power law $x^{(1+b)/b}$. We note that,
for instance, the power lay decay is present in the fractional and nonlinear
approaches \cite{klafter,pedron}.

For $b=1$ and $c=1/2$ we obtain

\begin{equation}
W(x,t)=C_{3}\frac{\left[ 2+a\sqrt{\left| x\right| }\right] \exp \left[ -%
\frac{x^{2}}{4D\overline{t}\left( 1+a\sqrt{\left| x\right| }\right) ^{2}}%
\right] }{\sqrt{\overline{t}}\left( 1+a\sqrt{\left| x\right| }\right) ^{2}}
\text{  .}\label{eq11}
\end{equation}
In this process, the behavior, for small $a$ and $x$, is like a Gaussian
function. For large distance, we have a exponential decay basically. We
note that the exponential decay has been observed in pair dispersion in
two-dimensional turbulence \cite{jullien}.

Moreover, for $bc<1/2$, the decay of the solution (\ref{eq9}) is essentially 
compressed Gaussian shape, whereas for $1/2<bc<1$, the decay is essentially 
stretched Gaussian shape. It is interesting to emphasize that the solution 
(\ref{eq9}) can have a similar asymptotic non-Gaussian shape of the random
walk model and time-fractional dynamic equation \cite{klafter}. The 
asymptotic shape of the random walk model and time-fractional dynamic 
equation is given by
 
\begin{equation}
W(x,t)\sim C_{4}t^{-\frac{\alpha}{2}}{\xi}^{-\frac{1-\alpha}{2-\alpha}}
 \exp \left[ -C_{5}{\xi}^{\frac{2}{2-\alpha}} \right]
\text{ ,}  \label{eq99}
\end{equation}
where $\xi \equiv \left| x\right| / t^{\alpha /2}$. This shape can be 
obtained from the solution (\ref{eq9}), for large distance, by taking 
$bc=(1-\alpha )/(2-\alpha )$, $\overline{t}=t^{\alpha /(2-\alpha )}$ and 
$T^{2}(t)=\frac{\alpha}{2-\alpha} t^{2(\alpha -1)/(2-\alpha)}$.

Second case - We consider

\begin{equation}
D(x)=\sqrt{D}\left| x\right| ^{-\frac{\theta }{2}}\text{ ,}  \label{eq12}
\end{equation}
where $\theta $ is a real parameter. We should note that the diffusion
coefficient (\ref{eq12}) has been used to describe the diffusive process on
a fractal \cite{proca}. The probability distribution (\ref{eq4}) for the
spatial multiplicative noise term (\ref{eq12}) is given by

\begin{equation}
W(x,t)=\frac{\left| x\right| ^{\frac{\theta }{2}}\exp \left[ -\frac{\left|
x\right| ^{2+\theta }}{D(2+\theta )^{2}\overline{t}}\right] }{\sqrt{4\pi D%
\overline{t}}}\text{ .}  \label{eq13}
\end{equation}
In this process, we have the bimodal states. In fact, we can construct many
bimodal states by choosing different functions for $D(x)$. For $\theta =0$
we recover the Wiener process. Basically, for large distance, the
probability distribution (\ref{eq13}) has a non-Gaussian decay. The second
moment related to this process is given by

\begin{equation}
\left\langle x^{2}\right\rangle =\frac{\left[ D^{2}(2+\theta )^{4}\right] ^{%
\frac{1}{2+\theta }}\Gamma \left[ \frac{6+\theta }{2(2+\theta )}\right] 
\overline{t}^{\frac{2}{2+\theta }}}{\sqrt{\pi }}\text{ \ .}  \label{eq14}
\end{equation}
The solution (\ref{eq13}) also reproduces the asymptotic shape (\ref{eq99})
by taking $2+\theta =2/(2-\alpha )$, 
$\overline{t}=t^{\alpha /(2-\alpha )}$ and $T^{2}(t)=\frac{\alpha}{2-\alpha} 
t^{2(\alpha -1)/(2-\alpha)}$. 
 The second moment (\ref{eq14}) 
yields  $\left\langle x^{2}\right\rangle \sim t^{\alpha}$ which corresponds 
to the same behavior of the time-fractional diffusion equation \cite{klafter}. For this process, 
the multiplicative noise term corresponds to  $(g(x,t)\sim 
\left| x\right|^{a}t^{b})$  which has the same form
suggested by Hentshel and Procaccia to study the turbulent system \cite{richar}.

We can now compare with the solution obtained by the Ito approach for the
same Langevin equation using (\ref{eq12}). The solution has been obtained in 
\cite{kwok}, and for $T(t)=1$ it is given by

\begin{equation}
W_{I}(x,t)\sim \frac{\left| x\right| ^{\theta }\exp \left[ -\frac{\left|
x\right| ^{2+\theta }}{D(2+\theta )^{2}t}\right] }{t^{\frac{1+\theta }{%
2+\theta }}}\text{ .}  \label{eq15}
\end{equation}
The second moment yields

\begin{equation}
\left\langle x^{2}\right\rangle _{I}\sim t^{\frac{2}{2+\theta }}\text{ .}
\label{eq16}
\end{equation}
We see that these two approaches give different behaviors for the
probability distribution due to the multiplicative factors. However, the
second moment of these two approaches give the same behavior. Further, both
the distributions present the bimodal states. It is also interesting to
compare with an other approach which uses the postpoint discretization rule 
\cite{hanggi,kwok}. For this last case, the probability distribution does
not present the bimodal states, however, its second moment has the same power
law behavior of the Ito and Stratonovich approaches. We see that three
different approaches give different behaviors, but they give the same power
law behavior for the second moment.

In summary, we have investigated the usual one-dimensional Fokker-Planck
equation with variable diffusion coefficient in the Stratonovich approach.
We have considered a very general class of the multiplicative noise term, $%
g(x,t)=D(x)T(t)$, and we have presented the formal solution for the
probability distribution. Using the formal solution, we have analyzed some
particular solutions by choosing simple functions for $D(x).$ We have shown
interesting behaviors for the probability distribution such as non-Gaussian,
exponential and power law decays for large distance. As we can note that the
usual Fokker-Planck equation can describe many different anomalous processes
with many different behaviors. The introduction of a time dependent multiplicative noise
term  may be necessary for the cases of more complex systems such as
turbulent systems, as suggested by several authors \cite{richar}. In fact,
we have shown that the asymptotic shape of the random walk model and 
time-fractional dynamic equation can be obtained from the solutions 
described in this work with the time dependent multiplicative noise term.
 Further, we have also shown that the solutions of the Ito, Stratonovich 
and postpoint discretization approaches, for $D(x)=\sqrt{D}\left| x\right| ^{-\theta /2}$,
describe different behaviors, but their second moments describe the same
behavior. If the diffusion coefficient $D(x)=\sqrt{D}\left| x\right|
^{-\theta /2}$ may describe exactly a real physical system by the Fokker-Planck
equation, then further information of the microscopic structure of the
system is necessary in order to choose which of the above approaches is the
correct one or, simply, which of the above approaches can fit the
experimental data.

\end{document}